\documentstyle[prl,aps,twocolumn,epsf]{revtex}

\begin{document}
\draft
\twocolumn[
\hsize\textwidth\columnwidth\hsize\csname @twocolumnfalse\endcsname

\title{Tunneling transverse to a magnetic field, \\and how it occurs
in correlated 2D electron systems}

\author{T. Barabash-Sharpee$^{(a)}$, M.I.~Dykman$^{(a)}$, and P.M.~Platzman$^{(b)}$}
\address{$^{(a)}$Department of Physics and Astronomy, 
Michigan State University, East Lansing, Michigan 48824\\
$^{(b)}$Bell Laboratories, Lucent Technologies, Murray Hill, New Jersey 07974}
\date{\today} 
\maketitle
\widetext 
\begin{quote}

We investigate tunneling decay in a magnetic field. Because of broken
time-reversal symmetry, the standard WKB technique does not apply.
The decay rate and the outcoming wave packet are found from the
analysis of the set of the particle Hamiltonian trajectories and its
singularities in complex space.  The results are applied to
tunneling from a strongly correlated 2D electron system in a magnetic
field parallel to the layer. We show in a simple model that electron
correlations exponentially strongly affect the tunneling rate.
\end{quote} 
] 
\narrowtext

Tunneling plays a fundamental role in many physical phenomena. In the
last decades much progress has been made in describing it beyond
the one-dimensional approximation and understanding how it occurs in
many-body systems
\cite{Langer,Caldeira,Kagan,AK}.  
For charged particles, the tunneling rate can be conveniently
controlled by a magnetic field applied transverse to the tunneling
direction.  Recently this effect was used to probe two-dimensional
electron systems (2DES) in semiconductor heterostructures
\cite{Smoliner,Eisenstein,MacDonald,minigaps} and 
on helium surfaces \cite{Andreimagn}. However, despite its interest
and generality, even the problem of single-electron tunneling decay in
a magnetic field (TDMF) remains largely unsolved. Existing results,
although often highly nontrivial, are limited to the cases where the
potential has either a special form \cite{Caldeira,Fert_Halperin,Ao}
(e.g., parabolic \cite{Fert_Halperin}), or a part of the potential or
the magnetic field are in some sense weak
\cite{Shklovskii,Thouless,Feng,Hajdu,Helffer}. 

Of particular interest  for the present paper is tunneling transverse to
the field from strongly correlated 2DES \cite{Andreimagn,Tsui}. In such systems
exchange is small, and the tunneling electron can be thought of as
moving in the potential created by other electrons. This motion can
exponentially strongly affect the tunneling rate. This can be
qualitatively understood, because an electron which tunnels a distance
$z$ transverse to the field $B$ and is free to move in the ${\bf
B}\times \hat{\bf z}$ direction, has its velocity $v_{_H}$ in this
direction incremented by $\omega_cz$ ($\omega_c=eB/m$ is the cyclotron
frequency). If initially $v_{_H}=0$, the energy of motion in the
$\hat{\bf z}$ direction is reduced by $m\omega_c^2z^2/2$, i.e. there
arises a parabolic magnetic barrier for tunneling.  On the other hand,
if an electron can give the momentum in the ${\bf B}\times \hat{\bf z}$
direction to in-plane excitations in the electron system, $v_{_H}$ can
remain small, effectively reducing this barrier.  This makes it possible
to use tunneling in a magnetic field as a sensitive probe of in-plane
electron dynamics in correlated systems. 

In this paper we will use an Einstein model in which the in-plane
electron motion is a harmonic vibration about an equilibrium position,
with one frequency (see Fig.~2 below) \cite{to_be}. The problem is then
effectively reduced to a single-particle problem, which mimics the
many-electron one. As we show, the resulting  tunneling exponent depends
on the dimensionless parameters $\omega_c\tau_0$ and $\omega_0\tau_0$,
where $\tau_0$ is the  imaginary time of underbarrier motion for $B=0$, 
 and $\omega_0$ is a
characteristic frequency of in-plane electron vibrations.

For smooth potentials and magnetic fields, the tunneling rate can be
found in the WKB approximation, in which the wave function is

\begin{equation}
\label{WKB1}
\psi({\bf r}) =D({\bf r})\exp[iS({\bf r})]\quad (\hbar =1).
\end{equation}

\noindent
Here,  $S({\bf r})$ is the classical action. It is calculated using the
classical equations of motion 

\begin{eqnarray}
\label{eom}
\dot S={\bf p}\cdot\dot{\bf r}, \quad \dot{\bf r} = \partial
H/\partial {\bf p},\;  \dot{\bf p} = -\partial
H/\partial {\bf r},
\end{eqnarray}

\noindent
where $H=({\bf p}+e{\bf A})^2/2m + U({\bf r})$ is the electron
Hamiltonian, and ${\bf A}$ is the vector potential of the magnetic field. 

In the standard approach to tunneling decay, which applies for $B=0$
\cite{Langer,AK,LandauQM,Schmid}, the action $S$ is purely imaginary
under the barrier. It is calculated by changing to imaginary time and
momentum in Eqs.~(\ref{eom}), which then take the form of equations of
classical motion in an inverted potential $-U({\bf r})$, with energy
$-E\geq -U({\bf r})$. In the presence of a magnetic field, because of
broken time-reversal symmetry, the replacement $t\to -it, \, {\bf p}
\to i{\bf p}, \, {\bf r}\to {\bf r}, \, U({\bf r}) \to -U({\bf r})$
would lead to a complex Hamiltonian, which makes no sense and
indicates that a more general approach is required.

We will find the action $S$ by solving the Hamiltonian equations
(\ref{eom}) in complex time and phase space. In contrast to the
$B=0$-case, in the presence of a magnetic field the action is {\it
complex} for {\it real} ${\bf r}$, i.e. the decay of the wave function
(\ref{WKB1}) under the barrier is accompanied by spatial oscillations.
The tunneling rate is determined by Im~$S$ at the point where the
particle emerges from the barrier as a semiclassical wave packet, with
real coordinate and real momentum.  However, again in contrast to the
standard ($B=0$) analysis, at this point the particle velocity is {\it
finite}, Re~$\dot{\bf r}\neq 0$.

The trajectories (\ref{eom}) of interest for tunneling decay start for
$t=0$ from the vicinity of the localized metastable state. The initial
conditions can be obtained from the known form of the wave function
$\psi({\bf r})$ close to the potential well, both in the case where
$U({\bf r})$ is parabolic near the minimum and $\psi$ is Gaussian
\cite{Langer,Caldeira,AK}, and where $U({\bf r})$ is nonanalytic in
one variable ($z$), which is of interest for 2DES. In both cases the
trajectories (\ref{eom}) are parametrized by two complex
parameters $x_{1,2}(0)$, which  can be the initial
values of the in-plane coordinates $x(0)= x_1(0),\, y(0)=
x_2(0)$ for given $z(0)$, and which in turn determine 
${\bf p}(0)$ and  $S(0)$, cf.  Eq.~(\ref{initial}).

To find the tunneling exponent we note that, once the particle has
escaped, it is described by a wave beam which propagates in real time
along a {\it real} classical trajectory ${\bf r}_{\rm cl}(t)$. This
trajectory can be obtained by analyzing the fan of complex
trajectories ${\bf r}(t), {\bf p}(t)$ (\ref{eom}) for different
$x_{1,2}(0)$ and finding such $x_{1,2}(0)$ that, for some $t$, both
${\bf r}(t)$ and ${\bf p}(t)$ become {\it real},

\begin{equation}
\label{exit}
{\rm Im}~{\bf r}(t) = {\rm Im}~{\bf p}(t) = 0.
\end{equation}

\noindent
This is a set of equations for complex $x_{1,2}(0)$ and Im~$t$. The
number of equations is equal to the number of variables, with account
taken of $H$ being real. The Re~$t$ remains undetermined: a change in
Re~$t$ in (\ref{exit}) results just in a shift of the particle along
the classical trajectory ${\bf r}_{\rm cl}(t)$, see Fig.~1. Such a shift
does not change Im~$S$. We note that, in contrast to what happens for
$B=0$, the classical trajectory does {\it not} have to touch the
boundary of the classically forbidden region.

The tunneling exponent
${\cal R}$ is given by the value of Im~$S$ at any point on the
trajectory ${\bf r}_{\rm cl}$,

\begin{equation}
\label{answer}
{\cal R} = 2~{\rm Im}~S({\bf r}_{\rm cl}).
\end{equation}

\begin{figure}
\begin{center}
\epsfxsize=3.0in                %so many inches wide
\leavevmode\epsfbox{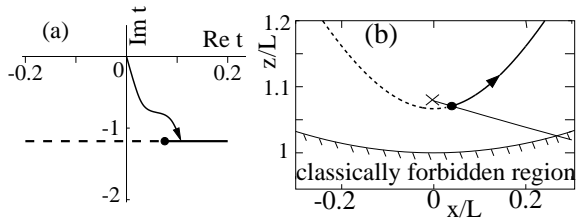}
\end{center}
\caption{(a) Complex $t$ plane for integrating the Hamiltonian equations
(\ref{eom}) in the escape problem. The line Im~$t$ = const corresponds
to the classical trajectory of outgoing electron.  (b) The classical
trajectory ${\bf r}_{\rm cl}$ on the $(x,z)$ plane. The bold solid
lines in (a) and (b) show the ``visible'' part of the trajectory where
the amplitude of the outgoing wave packet exceeds the amplitude of the
tail of the underbarrier wave function.  The thin solid line in (b)
shows where the amplitudes of the two wave functions coincide. It
starts from the caustic ($\times$). The data refer to the potential
(\ref{U(z)}), with $\omega_0\tau_0=1.2$ and $\omega_c\tau_0=1.2$, time
in (a) is in the units of $\tau_0$.}
\end{figure}
\noindent
For a physically meaningful solution, Im~$S$ has a 
parabolic minimum at ${\bf r}_{\rm cl}$ as a function of the coordinates transverse to the
trajectory, and the outgoing beam is Gaussian near the maximum. 

From (\ref{exit}), the tunneling exponent can be obtained by solving
the equations of motion (\ref{eom}) in imaginary time, with complex
${\bf r}$. However, such solution does not give the wave function for
real ${\bf r}$ between the well and the classical trajectory ${\bf
r}_{\rm cl}$. Neither does it tell us where the particle shows up on the
classical trajectory.  To obtain a complete solution of the tunneling
problem, one should take into account the fact that $S$ is a {\it
multivalued} function of ${\bf r}$, even though it is a single-valued
function of $t$ and $x_{1,2}(0)$, i.e. several trajectories
(\ref{eom}) with different $t$ and $x_{1,2}(0)$ can go through one and
the same point ${\bf r}$. The wave function is determined generally by
one of the branches of $S({\bf r})$.

Branching and multivaluedness of the action  are familiar
from 1D tunneling problem, where $S-S_t\propto (z-z_t)^{3/2}$ near the
turning point $z_t$ \cite{LandauQM}. In multidimensional systems,
branching generally occurs on caustics
\cite{Berry}. In our problem, in contrast to the usually considered
case, the trajectories ${\bf r}(t)$ will be {\it complex}, as will
also be the caustics.  Caustics of most general type are {\it
envelopes} of the trajectories ${\bf r}(t)$ (\ref{eom}). They are
given by the equation

\begin{equation}
\label{caustic}
J({\bf r})=0,\; J({\bf r})={\partial( x,y,z)\over \partial
(x_1(0),x_2(0),t)}.
\end{equation}

\noindent
The prefactor in the WKB wave
function (\ref{WKB1}) is $D={\rm const}\times J^{-1/2}$.  Therefore
the WKB approximation does not apply close to the caustic (cf. \cite{Berry}).

The caustic of interest is the one where the analytically continued
wave functions of the outgoing semiclassical wave and the WKB tail of
the intrawell state are connected.  The amplitude of the semiclassical
wave incident on the barrier from $z\to\infty$ should be set equal to
zero.  Local analysis near the caustic is similar to
that in the 1D case. It is convenient to change to the variables
$x',~y'$, and $z'$ which are locally parallel and perpendicular to the
caustic surface, respectively. We set $z'=0$ on the caustic. 
For small $|z'|$, 

\begin{eqnarray}
\label{increment}
S(x',y',z')\approx S(x',y',0) + a_1z' + a_2 z'^{3/2}
\end{eqnarray}

\noindent
(the coefficients $a_{1,2}\equiv a_{1,2}(x',y')$ can be expressed in
terms of the derivatives of $S,{\bf r}$ over $x_{1,2}(0),t$ on the
caustic). As in 1D problem, the prefactor in (\ref{WKB1}) depends on
the distance to the caustic as D$\propto (z')^{-1/4}$. However, in the
present case $S$ contains a linear term $a_1z'$, and therefore the
momentum perpendicular to the caustic is {\it finite}.  We note that
Eq.~(\ref{caustic}) and the condition Im~${\bf r} = 0$ define a line
in real space, which can be called a caustic line.

For real ${\bf r}$, there is a {\it switching surface} which separates
the ranges where Im~$S({\bf r})$ is smaller for one or the other of
the solutions connected on the caustic \cite{DMS}. Only the solution with the
smaller Im~$S$ should be held in the WKB approximation. It is this
condition that determines {\it where} the outgoing wave shows up from
beneath the tail of the intrawell state. The switching surface starts
from the caustic line, where the branches of $S$ merge together. The
cross sections of the switching surface and the caustic line by the
plane $(x,z)$ are shown in Fig.~1.

We now apply these general results to a simple model which is relevant
to electrons on helium. The corresponding geometry is shown in Fig.~2a.
For typical densities $n\sim 10^8$~cm$^{-2}$ and temperatures
T~$\alt$~1~K, these electrons form a Wigner crystal \cite{Wigner} or a
nondegenerate liquid \cite{PRL93}. In both these cases, the
characteristic frequency of vibrations about a (quasi)equilibrium
in-plane position is $\omega_0 =(2\pi e^2n^{3/2}/m)^{1/2}$.

The initial conditions for the equations of underbarrier motion
(\ref{eom}) can be chosen at an arbitrary plane $z=$ const close to
the electron layer and yet deep enough under the barrier so that the
electron wave function is semiclassical. We set $z\equiv z(0)=0$ on
this plane. In the Einstein model, the potential for in-plane motion
for $z=0$ is $(m\omega_0^2/2)(x^2+y^2)$, and $\psi(x,y,z=0)$ is
Gaussian in $x,y$.  Then
for $t=0$ in
(\ref{eom}),

\begin{eqnarray}  
\label{initial}  
z(0)=0,\; p_z(0)=i\gamma,\;
S(0)={im\omega_0\over 2}\left[x^2(0)+ y^2(0)\right],  
\end{eqnarray}

\noindent where  $\gamma = [2mU({\bf r}={\bf 0})]^{1/2}$ is the
reciprocal localization length in the $z$-direction, $\gamma\gg
(m\omega_0)^{1/2}$ (the energy of the localized state is set equal to
zero). The initial values of the in-plane momentum components $p_j(0)=
\partial S(0)/\partial x_j(0) = im\omega_0 x_j(0)$ (where $j=1,2$
enumerates the in-plane coordinates).

For electrons on helium, tunneling occurs if there is applied an
electric field ${\cal E}_{\perp}$ which pulls electrons away from the
helium surface \cite{AP}. The tunneling barrier is formed by the image
potential and the potential of this field. It is nearly triangular at
distances from the surface which are much greater than the effective
Bohr radius $\gamma^{-1}$. The barrier width $L=
\gamma^2/2me{\cal E}_{\perp}$ for $B=0$. Typically $L\ll n^{-1/2}$, and
therefore the effective electron potential energy is well represented by

\begin{equation} 
\label{U(z)} 
U({\bf r})= {m\omega_0^2\over 2}(x^2+y^2)
+ {\gamma^2\over 2m}\left(1-{z\over L}\right) \; (z>0). 
\end{equation}

With (\ref{U(z)}), the equations of motion (\ref{eom}) become linear
and can be readily solved.  The symmetry of the potential
$U(x,y,z)=U(\pm x,\pm y,z)$ gives rise to a specific symmetry of the
set of the trajectories (\ref{eom}), and one can show that the caustic
of interest intersects the real space for $x=0$ and some $z=z_c$
($z_c=L$ for $B=0$).

For ${\bf B}$ along the $y$-axis (see Fig.~2a), the motion in the $y$
direction is decoupled and the problem becomes two-dimensional. For
$z\leq z_c$, the function Im~$S$ has two branches each of which is
symmetrical in $x$.  The branch 1 describes the tail of the intrawell
wave function before branching. It has a minimum at $x=0$ for given
$z$, and monotonically increases with $|x|$ and $z$. As expected, the
slope $\partial~{\rm Im}~S/\partial z$ is finite at the branching
point $z_c$.  The branch 2 corresponds to the wave ``reflected'' from
the caustic.  This branch is nonmonotonic in $z$ for $x=0$,
\vspace*{-0.05in}
\begin{figure}
\begin{center}
\epsfxsize=2.9in                %so many inches wide
\leavevmode\epsfbox{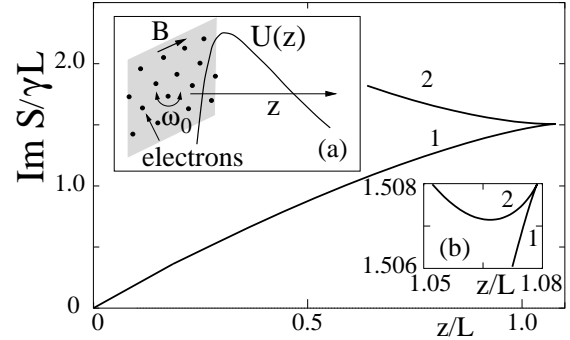}
\end{center}
\caption{Two branches of the action  on 
the symmetry axis $x=0$ as a function of the tunneling coordinate $z$
before the branching point, for the same parameters as in Fig.~1. The
vicinity of the cusp at $z_c$ is zoomed in inset (b).
Inset (a): the geometry of tunneling
from a correlated 2DES transverse to a magnetic field; $\omega_0$ is
the Einstein vibration frequency in the Wigner solid. }
\end{figure}
\vspace*{-0.12in}
\noindent
with a minimum at $z_m<z_c$. For $z_m < z\leq z_c$, it has two symmetrical
{\it minima} for $x\neq 0$. They lie on the classical trajectory shown
in Fig.~1b, and merge together for $z=z_m$.

As discussed above, Im~$S$ is constant on the classical trajectory in
Fig.~1b. This trajectory goes through the point $x=0, z=z_m$ and is
symmetrical in $x$.  Although the potential $U({\bf r})$ is even in
$x$ and is minimal for $x=0$, the escaped particle ``shows up'' on the
classical trajectory for finite $x$. This happens where Im~$S_1({\bf
r}_{\rm cl}) =$ Im~$S_2({\bf r}_{\rm cl}) ={\cal R}/2$ (the subscript
enumerates the branches in Fig.~2). The particle has finite velocity
and moves away from the barrier.

Since the point $x=0,\, z=z_m$ lies on the classical trajectory of
interest (although on the section ``hidden'' by the tail of the
intrawell state), the tunneling exponent is given by ${\cal R}= 2\,{\rm
Im}\,S_2(x=y=0,z_m)$ and can be calculated in imaginary time,
with imaginary $x(0)$

\begin{eqnarray}
\label{R}
\tilde{\cal R}&& = -\nu_0^2\tau^3-
3\nu_0(1-\tau)^2 + 3(1+\nu_0)(1-\tau)\\ 
&&+3\nu^2\tau,\quad {\cal R}=2\gamma L\tilde{\cal R}/3\nu^2,
\nonumber
\end{eqnarray}

\noindent
where $\nu_0=\omega_0\tau_0, \nu_c=\omega_c\tau_0$ are the
dimensionless in-plane and cyclotron frequencies ($\tau_0=2mL/\gamma$
is the ``duration'' of underbarrier motion in imaginary time for $B=0$),
$\nu^2=\nu_0^2+\nu_c^2$, and $\tau =it/\tau_0$ is given by the equation

\begin{eqnarray}
\label{tau}
\left[\nu^2\nu_0(1-\tau) -\nu_c^2\right]\tanh\nu\tau = \nu(\nu_0^2\tau-\nu^2).
\end{eqnarray}

The tunneling exponent as a function of $\omega_0,\omega_c$ is shown
in Fig.~3. For
$\omega_0=0$ (no electron-electron interaction), the magnetic barrier
makes tunneling impossible for $\omega_c\tau_0\geq 1$
\cite{Andreimagn} (see curve 1; $\tau\to\infty$ for $\nu_0=0,\nu_c\to
1$). Even comparatively weak in-plane confinement eliminates this
effect. The reduction of the tunneling suppression is significant
already for small $\omega_0\tau_0$, and increases fast with increasing
$\omega_0\tau_0$.

\begin{figure}
\begin{center}
\epsfxsize=2.65in                %so many inches wide
\leavevmode\epsfbox{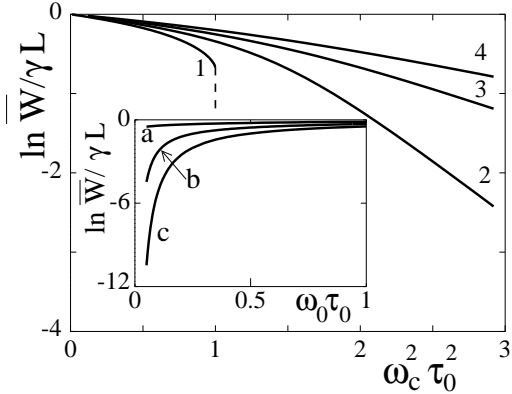}
\end{center}
\vspace*{-0.1in}
\caption{The dependence of the tunneling rate on magnetic field, 
$\bar W = W(B)/W(0)$. The curves 1 to 4 refer to $\omega_0\tau_0 =
0,0.2,0.4,0.6$.  Magnetic field eliminates single-electron tunneling
for $\omega_c\tau_0\geq 1$ (cf. curve 1). Inset: tunneling exponent vs
in-plane frequency $\omega_0$ for $\omega_c^2\tau_0^2 = 1.0,2.0,3.0$
(curves a,b,c).}
\end{figure}
The above results provide an explanation of the magnetic field
dependence of the tunneling exponent for electrons on helium, which
was observed to be {\it much weaker} \cite{Andreimagn} than it would
be expected from the single-electron theory. Detailed comparison with
the data \cite{Andreimagn} will be discussed elsewhere \cite{to_be},
where the model will also be extended in order to include the
realistic vibrational spectrum of the Wigner solid. At zero
temperature this extension does not change the results significantly,
because electron tunneling is accompanied by excitation of mostly
short-wavelength vibrations,
which are reasonably well described by the Einstein model used above.

In conclusion, we have shown that, under suitable conditions
($\omega_c\tau_0\agt 1,\,\omega_0\tau_0\agt 1$), correlations in a
2DES can exponentially strongly affect the rate of tunneling escape
transverse to a magnetic field. We have also shown that the problem of
single particle tunneling in a magnetic field can be solved in the
semiclassical limit by analyzing the Hamiltonian trajectories of the
particle in complex space and time. The connection of decaying and
propagating waves occurs on caustics of the set of these trajectories.
This approach does not require us to consider any piece of the
electron potential or the magnetic field as a perturbation. It gives
us an escape rate which is generally {\it exponentially} smaller than
the probability for a particle to reach the boundary of the
classically accessible range $U({\bf r})= E$. Finally, we have
obtained explicit results for a simple model of an electron tunneling
from a helium surface transverse to a magnetic field.

We are grateful to V.N. Smelyanskiy who participated in this work at
the early stage, and to D. Farber for help with numerical
calculations. This research was supported in part by the NSF through
Grant no. PHY-9722057.


\begin{thebibliography}{99} 

\bibitem{Langer} J.S. Langer, Ann. Phys. {\bf 41}, 108 (1967);
S. Coleman, Phys. Rev. D {\bf 15}, 2929 (1977).


\bibitem{Caldeira} A.O. Caldeira and A.J. Leggett, 
Ann. Phys. {\bf 149}, 374 (1983).

\bibitem{Kagan} 
{\it Quantum Tunnelling in Condensed Matter}, eds. Yu. Kagan and
A.J. Leggett (Elsevier, NY 1992).

\bibitem{AK} A. Auerbach and S. Kivelson, Nucl. Phys. {\bf B257}, 799 (1985).

\bibitem{Smoliner} J. Smoliner {\it et al.}, Phys. Rev. Lett. {\bf
63}, 2116 (1989); G. Rainer {\it et al.}, Phys. Rev. B {\bf 51}, 17642
(1995).

\bibitem{Eisenstein} J.P. Eisenstein {\it et al.}, 
Phys.  Rev. B {\bf 44}, 6511 (1991); S.Q. Murphy {\it et al.},
Phys.  Rev. B {\bf 52}, 14825 (1995).

\bibitem{MacDonald} L. Zheng and A.H. MacDonald,
Phys. Rev. B {\bf 47}, 10619 (1993).

\bibitem{minigaps} T. Ihn {\it et al}, Phys. Rev. B {\bf 54}, R2315
(1996); M.J. Yang {\it et al.}, 
Phys. Rev. Lett. {\bf 78}, 4613 (1997); M. Lakrimi {\it et al.},
Phys. Rev. Lett. {\bf 79}, 3034 (1997).

\bibitem{Andreimagn} L. Menna, S. Y\"{u}cel, and  E.Y. Andrei, Phys. Rev.
Lett.  {\bf 70}, 2154 (1993).

\bibitem{Fert_Halperin} H.A.~Fertig and B.I.~Halperin, 
Phys. Rev. B {\bf 36}, 7969 (1987).

\bibitem{Ao} P. Ao,  Phys. Rev. Lett. {\bf 72}, 1898
(1994); Phys. Scripta {\bf T69}, 7 (1997).

\bibitem{Shklovskii} B.I. Shklovskii, JETP Lett. {\bf 36}, 51 (1982);
B.I. Shklovskii and A.L. Efros, Sov. Phys. JETP {\bf 57}, 470 (1983).

\bibitem{Thouless} Qin Li and D.J.~Thouless, Phys. Rev. B {\bf 40},
9738 (1989).

\bibitem{Feng} T.~Martin and S.~Feng, Phys. Rev. B {\bf 44}, 9084 (1991).

\bibitem{Hajdu} J.~Haidu, M.E.~Raikh, and T.V.~Shahbazyan, 
Phys. Rev. B {\bf 50}, 17625 (1994).

\bibitem{Helffer} B.~Hellfer and J.~Sj\"{o}strand, 
Ann. Scuola Norm. Sup. Pisa Cl. Sci. (4) {\bf 14},
625 (1988).

\bibitem{Tsui} Jongsoo Yoon {\it et al.}, Phys. Rev. Lett. {\bf 82},
1744 (1999); A.P. Mills, Jr. {\it et al.}, Phys. Rev. Lett. {\bf 83},
2805 (1999), and references cited in these papers.

\bibitem{to_be} T. Sharpee {\it et al.},
in preparation.

\bibitem{LandauQM} L.D. Landau and E.M. Lifshitz, {\it Quantum
mechanics: non-relativistic theory} (Pergamon, NY 1977).
M.V. Berry and K.E. Mount, Rep. Progr. Phys. {\bf 35}, 315 (1972).

\bibitem{Schmid} U. Eckern and A. Schmid, in Ref.~\cite{Kagan}, p.~145.

\bibitem{Berry} M.V. Berry, Adv. Phys. {\bf 25}, 1 (1976);
L.S. Schulman, {\it Techniques and applications of path integration}
(Wiley, New York, 1981).

\bibitem{DMS} For classical systems, the ocurrence of switching 
lines on the tails of distribution was discussed by M.I. Dykman,
M.M. Millonas, and V.N. Smelyanskiy, Phys. Lett. A {\bf 195}, 53 (1994).

\bibitem{Wigner} C.C. Grimes and G. Adams, Phys. Rev.  Lett. {\bf 42},
795 (1979); D.S. Fisher, B.I. Halperin, and P.M. Platzman, Phys. Rev.
Lett. {\bf 42}, 798 (1979).

\bibitem{PRL93} M.I. Dykman {\it et al.}, Phys. Rev.  Lett. {\bf 70},
3975 (1993); M.J. Lea and M.I. Dykman, Physica B {\bf 251}, 628 (1998).

\bibitem{AP}% Tunneling of electrons on helium for $B=0$ was discussed by 
M.Ya. Azbel and P.M. Platzman, Phys. Rev. Lett. {\bf 65}, 1376 (1990).

\end{thebibliography}
\end{document}